\documentclass[prl,twocolumn,showpacs]{revtex4}
\usepackage{epsfig}
\usepackage{amsbsy}
\usepackage{amsmath}
\usepackage{graphicx}
\usepackage{latexsym}

\begin{document}

\title{Localized Entanglement in one-dimensional Anderson model}
\author{Haibin Li$^{(1)}$\footnote{
Electronic address: hbli@zjut.edu.cn}, XiaoGuang Wang$^{(2,3)}$}

\affiliation{1. Department of applied physics, Zhejiang University
of Technology, Hangzhou 310014, China.}

\affiliation{2. Zhejiang Institute of Modern Physics, Zhejiang
University, Hangzhou 310027, China.}

\affiliation{3. Department of Physics and Australian Centre of
Excellence for Quantum Computer Technology,\\
Macquarie University,Sydney, New South Wales 2109, Australia.}

\date{\today}
\begin{abstract}
The entanglement in one-dimensional Anderson model is studied. The
pairwise entanglement has a direct relation to the localization
length and is reduced by disorder. Entanglement distribution
displays the entanglement localization. The pairwise entanglements
around localization center exhibit a maximum as the disorder
strength increases. The dynamics of entanglement is also
investigated.
\end{abstract}
 \maketitle Entanglement is a
kind of nonlocal correlation that only exists in quantum systems.
Recent studies on entanglement are motivated by its potential
applications in quantum computation, quantum teleportation and
quantum communication. As spin system is a perfect one to realize
quantum computer, many efforts focus on the entanglement in the
Heisenberg spin model~\cite{MA, MAa}, Ising model in a transverse
magnetic field~\cite{MA2,MA2a} and itinerant fermionic
systems~\cite{MA3}. In quantum computer, one needs to control and
measure on individual qubits. However, in many possible physical
implementations of quantum computer, the interaction between spin
and spin (or qubit-qubit) is inevitable, and excitation or
paticles can hop from one site to other sites. So it is hard to
operate on one single qubit. To overcome this difficulty, one
excitation should be pinned on a certain site. In fact, it is well
known that the localization can pin the excitation.

On the other hand,being a fundamental concept of quantum theory,
entanglement must be involved in many fields of physics. It is
shown that entanglement is a indicator of quantum phase
transition~\cite{MA4,MA4a,Cirac}. The relation between
entanglement and chaos, and the relation between entanglement and
localization are also discussed, and it was found that strong
localization decreases entanglement~\cite{San}.The quantum
entanglement in condensed matter system, bose system and fermi
system, and their connection with long-range order and spontaneous
symmetry breaking are also discussed~\cite{Yu1,Yu2}.

The effect of incommensurate coupling strength on pairwise
entanglement was studied by considering the one-particle states of
the Harper model~\cite{Arul} and Frenkel-Kontorova
model~\cite{XX}, respectively. The behaviors of entanglement
during the phase transition and relation between pairwise
entanglement and state localization were revealed. Moreover, there
is also a simple quantitative relation between the bipartite
entanglement and the state localization~\cite{HH}.

Disorder is a common factor which exists in a large area of
physical world. It is well-known that disorder can lead to
localization from the Anderson's early work~\cite{AD}, which
influences many properties of physical system such as the electric
insulator. In this paper, we study the effects of disorder on
entanglement of one-particle states in the one-dimensional
Anderson model. We show that the localization decreases the
entanglement sharing in one particle state and entanglement is
shown to be an indicator of the localization in one-dimensional
systems.

In general, one-particle state belongs to a subspace of the
$2^N$-dimensional Hilbert space of $N$ qubits. This subspace is
$N$-dimensional and spanned by states with only one excitation.
One particle state can be written as
\begin{eqnarray}
|\Psi \rangle  &=&\psi _1|1000\ldots 0\rangle +\psi _2|0100\ldots
0\rangle
\label{eq1} \\
&&+\ldots +\psi _N|0000\ldots 1\rangle.   \nonumber
\end{eqnarray}
this state can be realized in many quantum systems such as those
of one spinless fermion or boson hopping on a substrate, and one
magnon in Heisenberg spin systems.

For a pure state of bipartite system, composing of two subsystems
$A$ and $B,$ the bipartite entanglement can be measured by Von
Neumann entropy, linear entropy, or other entropies. For mixed
state of two qubits $\rho$,  Wootters et.al~\cite{WW} had found
that the entanglement of formation is monotonic function of its
concurrence which is defined by $ C(\rho )=\max (0,\lambda
_1-\lambda _2-\lambda _3-\lambda _4) \label{3} $ where $\lambda
_1,\lambda _2,\lambda _3,\lambda _4$ are the square roots of the
matrix $ \stackrel{\sim }{\rho }=\rho(\sigma _y\otimes \sigma
_y)\rho ^{*}(\sigma _y\otimes \sigma _y)$. For the general
one-particle state, the entanglement between a pair of qubits,
qubit $i$ and $j$, can be quantified by the concurrence given
by~\cite{Coffman}
\begin{equation}
C_{i,j}=2\mid \psi _i\psi _j\mid   \label{7}
\end{equation}

One electron hopping on substrate potential can be described by a
general Hamilton $H=\frac{P^2}{2m}+V$. If we are not interested in
the behavior of the wave function on length scales smaller than a
lattice constant, the model can be described by the discrete
Schr\"{o}dinger equationwith with a fixed lattice constant $a$
\begin{equation}
-\frac \hbar{2ma^2}(\psi _{i+1}-2\psi _i+\psi _{i-1})+V_i\psi
_i=E_i\psi _i, \label{8}
\end{equation}
which can be written in more comprehensive way as
\begin{equation}
-t(\psi _{i+1}+\psi _{i-1})+\varepsilon _i\psi _i=E_i\psi _i.
\label{9}
\end{equation}
In the second quantized picture, the Hamiltonian can be written
further as the following
\begin{equation}
H=-t\sum_{i=1}^N (c_i^{\dagger}c_{i+1}+h.c)+V_ic_i^{\dagger}c_i,
\label{10}
\end{equation}
where $t$ is the nearest-neighbor hopping integral, measuring the
probability for electron transfer from n-th site to its
nearest-neighbor sites, it is chosen to be unit throughout this
paper. $V_i$ is the on-site potential.
 $c_i^{\dagger}$ and $c_i$
are the creation and annihilation operators of i-th local
fermionic modes, satisfying the canonical anti-commutation
relation, $\left[ c_i,c_j\right] _{+}=0,\, \left[
c_i,c_j^{\dagger}\right]_+=\delta _{ij}.$ The general state of the
electron hopping in the one-dimensional lattice can be viewed as a
multiqubit state~(\ref{eq1}) in the occupation-number basis, and
thus the entanglement between two local fermionic modes can be
discussed~\cite{ZA}.In fact, the single electron model is
equivalent to one magnon state of the XY model which is described
by the Hamiltonian $H=-t\sum_{n=1}^N(\sigma _n^{+}\sigma
_{n+1}^{-} +\sigma _n^{-}\sigma_{n+1}^{+})+\sum_{n=1}^Nh_n\sigma
_n^z$. So the entanglement properties of these two models are the
same, we can also understand the effect of localization on the
entanglement of pair spins(qubits).

Consider the motion of electron in the one-dimensional Anderson
model, the on-site potential can written as $ V_i=V_0+\lambda
\epsilon _i,$  where $\epsilon _i$ is a Gaussian random variable,
satisfying $\left\langle \epsilon_i \epsilon_j \right\rangle
\propto \delta _{ij},\left\langle \epsilon_i\right\rangle =0$, and
$\lambda $ is the disorder strength. To study global entanglement
of the system and reveal the relation between entanglement and
state localization, we use the entanglement measure given by the
average concurrence~\cite{Arul}
\begin{equation}
\left\langle C\right\rangle =\frac 1M\sum_{i<j}C_{ij}=\frac
1M\left[ \left( \sum_{i=1}^N|\psi _i|\right) ^2-1\right]
\label{14}
\end{equation}
with $M=N(N-1)/2$.

As mentioned in the introduction, the localization caused by
disorder is typical and important in condensed matter physics. The
direct result of disorder is the localization of state, i.e., the
wave function of system, which exhibits a exponential spatial
decrease~\cite{Lee},
\begin{equation}
\psi _i\sim \psi _{i_0}\exp \left(-\frac{|x_i-x_{i_0}|}\xi \right),
\label{15}
\end{equation}
where $\xi $ is the localization length, $x_i$ is the site
coordinate of wave function, and $x_{i_0}$ is the center site of
localization.

In the continuous limit, the average concurrence~(\ref{14}) can be
written as an integral. Considering the fast exponential decrease
of wave function of localized state, the average concurrence can
be estimated as
\begin{equation}
\left\langle C\right\rangle  \sim \frac 1M\left\{
\Big[2\int_{x_{i_0}}^{x_{i_0}+\xi }\psi(x) dx\Big]^2-1\right\} \label{16}
=\frac 4M\xi ^2+A,
\end{equation}
where $A$ is the constant. It is evident that the average pairwise
entanglement has a direct relation to the localization length in
the one-dimensional disorder system. The localization length is
the most important indicator of localization especially for the
disorder system. When disorder becomes stronger, the localized
length become small, the state of system is more localized and
less entangled due to the above analytic result~(\ref{16}).

\begin{figure}
\includegraphics[width=0.40\textwidth]{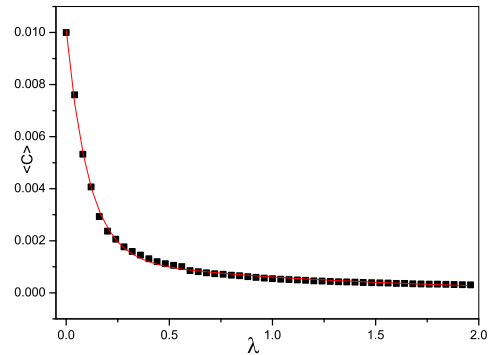}
\caption{Average concurrence as a function of disorder strength
parameter $\lambda $. Solid line is the fit curve of exponential
decay. The system size $N=1600$ and the concurrence is averaged
over 50 disorder ensembles.}
\end{figure}

To show the above result in more detail, we study the entanglement
of ground state of the present one-dimensional disorder system by
numerical calculations. In Fig.~1, the behavior of average
concurrence of the ground state of Hamiltonian(5) versus disorder
strength is presented. When the disorder is absent, i.e., $\lambda
=0.0$, the average concurrence exhibits a maximal value. The
increase of disorder strength leads to the decrease of average
concurrence. It is shown that the decrease of concurrence fits the
second exponential order
\begin{equation} \left\langle C\right\rangle \sim
B_1e^{-\lambda /D_1}+B_2e^{-\lambda /D_2} \label{17}
\end{equation}
with $B_1=0.00872,D_1=0.10611,B_2=0.00139,$ and $D_2=1.2166$. This
numerical result indicates that the disorder has great effects on
quantum entanglement, agreeing with the analytic
result~(\ref{16}).In the extended case of Hamiltonian(5) , the
system has a maximal entanglement so that this quantum correlation
corresponds to an ideal electric conductor. When the disorder is
introduced, the entanglement becomes small, the system turns from
extended to localized state, and the conduction of system becomes
finite and tends to be zero, indicating an insulator and no
quantum correlation. If we consider the thermal conductivity
carried by one spin excitation, the result is the same.

\begin{figure}
\includegraphics[width=0.40\textwidth]{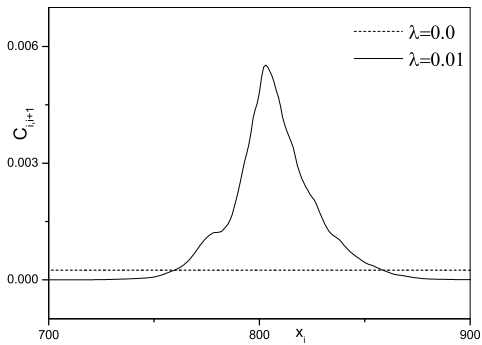}
\caption{The distribution of the nearest-neighbor pairwise
entanglement in the case of weak disorder.The system size
$N=1600$.}
\end{figure}

\begin{figure}
\includegraphics[width=0.40\textwidth]{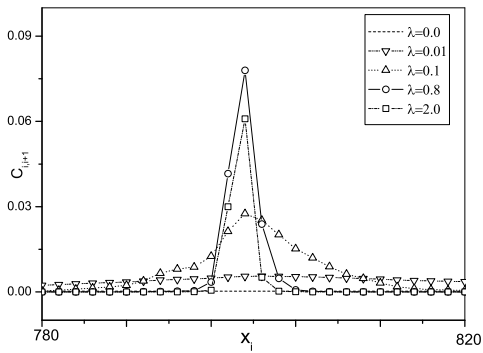}
\caption{The distribution of the nearest-neighbor pairwise
entanglement from weak disorder to strong disorder. The system
size $N=1600$.}
\end{figure}

In fact, the disorder leads to spatial exponential decrease of
eigenstates no matter how weak the disorder strength is~\cite
{MO}. When a localization of state occurs, what will be the
behaviors of pairwise entanglement? In the following, we will
discuss the entanglement between two nearest-neighbor sites, and a
distribution of the pairwise entanglement is expected. The
concurrence for the nearest-neighbor sites $i$ and $i+1$ is given
by $C_i=2\left| \psi _i\psi _{i+1}\right|$, and the numerical
results are given in Fig.~2.

In Fig.~2, we notice that the entanglement distributes among every
pair sites uniformly when the disorder is not present. Once the
disorder is introduced, the localization emerges and the
entanglement distribution becomes site-dependent. We can see that
even the strength of disorder is very weak, for example $\lambda
=0.01$, the entanglements between most of pairs are suppressed to
a very small value and only some pair of sites exhibit higher
pairwise entanglement around localization center. In other words,
the entanglement is constrained to some certain pairs of sites.
The distribution of pairwise entanglement with different values of
$\lambda$ are plotted in Fig.~3. When $\lambda $ increases, the
entanglements between more pairs of sites become close to zero,
and the number of pairs which have enhanced pairwise entanglement
becomes small. In other words, the width of localized entanglement
distribution becomes small and entanglement is constrained. We can
see that the localized entanglement is clear in the entanglement
distribution picture.

\begin{figure}
\includegraphics[width=0.40\textwidth]{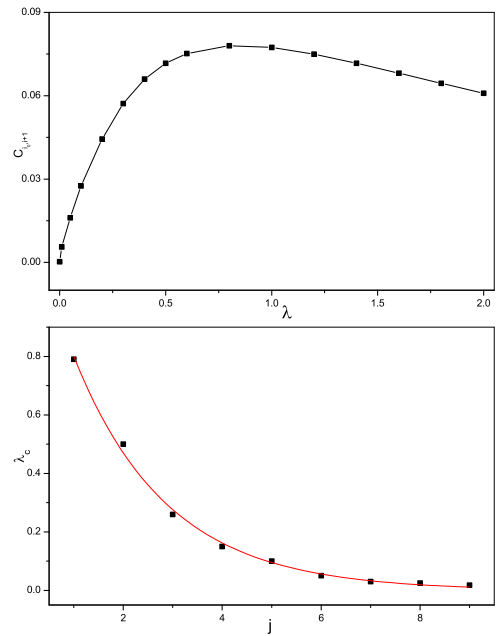}
\caption{(a)The maximal pairwise entanglement as a function of
disorder strength. $N=1600$ and $\gamma=0.1$.(b) The critical
value of $\lambda $ as a function of relative index $j$ which
label the distance from one site to the localization center site.}
\end{figure}

We also notice that the pairwise entanglement between localization
center site and its near sites are firstly enhanced in the weak
disorder regime and then suppressed in the strong disorder regime.
This property is shown in Fig.~4(a), the entanglement
$C_{i_{0}i_0+1}$ is plotted as the function of $\lambda $, where
$i_0$ denote the localization center site. There exists a maximal
value of entanglement at $\lambda =0.8.$  If $\lambda <0.8,$ the
entanglement is increasing. When $\lambda $ is beyond $0.8$, the
entanglement decreases linearly. Such behavior of the entanglement
can be understood as following. On one side, the increase of
$\lambda$ will lead to entanglement localization, thus enhance
$C_{i_{0}i_0+1}$. On the other side, from Hamiltonian~(\ref{10}),
we know that the increase of $\lambda$ will make the term $\sum_i
V_i c_i^\dagger c_i$ predominant, which suppresses entanglement in
the system. The competition between the two roles played by the
disorder strength leads to the maximum value of the concurrence
$C_{i_{0}i_0+1}$. We also study pairwise entanglements between
non-nearest-neighbor sites and the center sites. They also present
the maximal behavior during the increasing of disorder strength,
but the critical value of $\lambda $ will decrease when the site
is more far from the center site, which is shown is Fig.~4(b).
$\lambda_c $ is plotted as the function of the index $j$, where
$j$ is the relative index from the localization center and is
defined as $j=i-i_0$, $i_0$ is the index of localization center.
We note that the decrease of critical $\lambda $ is also
exponential, which coincide with the the exponential decay of
wavefunction.

\begin{figure}
\includegraphics[width=0.40\textwidth]{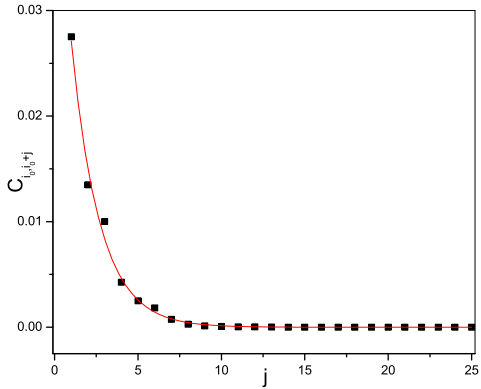}
\caption{The pairwise entanglement between the center site of
localization and $j$-th site, having distance
$l=\left|x_j-x_{i_0}\right|$ from center site. Solid line is the fit
curve of exponential decay. The size of the system $N=1600$.}
\end{figure}

We also consider the pairwise entanglement between the center site
and it's near sites when the disorder strength is fixed. In this
case, the concurrence is given by $ C_j=2\left| \psi
_{j+i_0}\psi_{i_0}\right|$, where the label $j$ is the relative
index from the localization center.As the amplitude of the wave
function has the maximal absolute value at $x_{i_0}$, by
considering Eq.~(7), $C_j$ becomes $2|\psi_{i_0}^2|\exp
(-|x_{j+i_0}-x_{i_0}|/\xi),$ which measures the entanglement
between the localization center and $j$-th site away from the
center. We calculate the entanglement and get the result with
different disorder strengths. As expected, the pairwise
entanglement decreases exponentially, agreeing with the simple
analytical result. This property is shown in Fig. 5.

\begin{figure}
\includegraphics[width=0.40\textwidth]{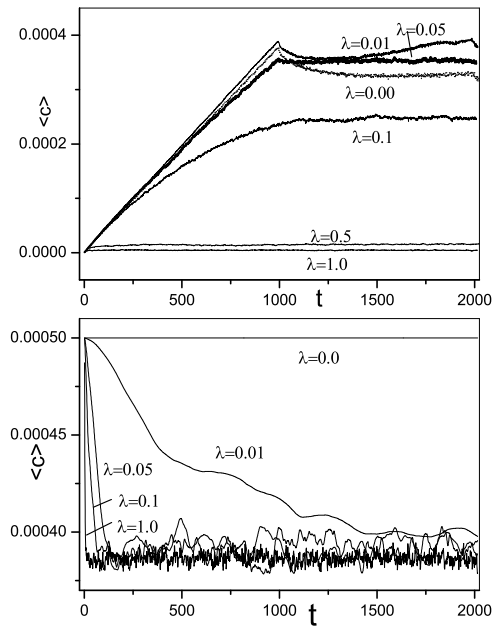}
\caption{Time evolution of the average concurrence of present
model with different disorder strength. The two extreme initial
condition is applied (a)Zero entanglement initial state. (b) The
maximal entanglement state.The system size $N=1600$.}
\end{figure}
From the above results of pairwise entanglement, we can find that
when disorder is absent, the pairwise entanglement between every
pair of sites is identical and independent on site index so that
the nonlocality  is completely shown. For spin system, one spin is
entangled with the other one far from it as much as the one near
it. This agrees with the result that in the extended state, the
system is a perfect conductor for electric current or spin
transport. If the strengthen of disorder is not zero, the pairwise
entanglement becomes localizable and dependent on site index. The
entanglement between nearest neighbor spins(sites) is different,
and spins(sites) will prefer to entangle with near spins (sites)
than the one far away. The nonlocality is destroyed, and
entanglement is decrease globally, although some local
entanglements between pair spins do not. In this case, one spin
excitation would be pinned at some sites and the conduction
vanishes.

We have studied the properties of ground-state entanglement in the
one-dimensional Anderson disorder model. Next, we will investigate
the dynamics of entanglement. This question can be solved by
calculating the time-dependent Schr\"{o}dinger equation
\begin{equation}
i\frac{d\psi _i}{dt}=-\psi _{i+1}-\psi _{i-1}+V_i\psi _i
\label{20}
\end{equation}
which can be integrated numerically.

The dynamical behavior of average concurrence of presenting model
are shown in Fig.~6 with different disorder strengths. Fig.6 (a)
displays the dynamical evolution of entanglement with the initial
state being $|\psi \rangle _0=|1\rangle
=c_{l}^{\dagger}|0\rangle,$ where $l$, indicating one site of
system. is set $l=N/2$ and $N$ is the size of system. In this
case, the initial state has no entanglement, what will this state
get entangled when time goes on? If disorder is absent, $\lambda
=0,$ the average concurrence increases linearly with time. When
time is long enough, the entanglement reaches a maximal value,
after that it will decrease to a smaller value. When the disorder
strength is not large, for example, $\lambda =0.05$, the
entanglement also increase linearly at first stage and reach a
maximal value. But the decrease of entanglement after maximal
value becomes small. We notice that when $\lambda =0.1,$ the
entanglement does not decrease, and almost keeps the maximal
value. When disorder strength is large enough, like $\lambda
=0.5$, the initial entanglement increases slowly. After reaching a
small value, it does not increase again. If $\lambda $ is
larger,for example $\left( \lambda =1.0\right),$ the entanglement
exhibits almost no increase.  This is a strong indicator of
localization.

If one start the Equ.~(\ref{20}) from the maximally pairwise
entangled initial state, namely the $W$ state~\cite{MAa,W}, the
dynamical behaviors are displayed in Fig 6.~(b). The maximal
entanglement is invariant while disorder is absent. But as soon as
the disorder is introduced, the entanglement decreases as time
goes on, the rate of the decrease is determined by the disorder
strength. The larger disorder strength will make entanglement
decrease quicker than smaller one. But it is interesting that no
matter how large the disorder strength is, the entanglements will
all decrease to a same residue value. Comparing the dynamical
evolution of the entanglements from two different initial states,
we find the asymptotic behaviors are different. The residue values
of initial maximal entangled state are larger than those of
initial non-entangled state for any value of disorder strength.

The evolution of entanglement in this system is determined by the
diffusion of wavefunction and the localization. Without disorder,
the wavefunction of electron will spread along the spatial
direction. The external disorder potential can localize the motion
of electron (or one excitation in XY spin system) on finite region
of space. The existence of these two process leads to the
dependence of evolution of entanglement on the initial condition
and disorder strength. For an initial unentangled state which has
a strong tendency to spread, a weak disorder can not prevent
occupation possibility of particle sharing among more sites, so
the entanglement will increase a lot, but a strong disorder will
make a localization of the state, which leads to a small increase
of entanglement. If the initial state is maximally entangled
state, a completely extended state, so only the localization will
determine the evolution entanglement. The entanglement must
decrease as time goes on and the disorder strength determines the
rate of decrease. We also do same calculations when we use other
initial states different from the above two states. The result is
also the same, namely, the disorder will destroy entanglement. By
studying these properties, we can find that other states, except
for ground state, have similar entanglement characters.So, it is
true that electric conduction with single electron approximation
and thermal conduction of one spin excitation are connected with
quantum correlation measured by pairwise entanglement.

In summary, we have studied the ground state and dynamical
pairwise entanglement of one-dimensional Anderson disorder model.
By simple analytic and numerical results, we can find the
entanglement is the indicator of localization caused by disorder.
The disorder can destroy such quantum correlation. The relations
between the entanglement and conduction are also discussed. On the
other hand, we can localize one qubit on certain site by disorder,
then we can do quantum operation on it.

It is interesting to consider entanglement in other disorder
models. These studies will strength our understanding of
entanglement, localization, and their relations. For instance, it
is an interesting question to study effects of disorder on
entanglement, the relation between localization and entanglement
for other subspace with the number of electrons or spin
excitations being large than one, which are under consideration.
In that model , the effect of disorder on multi-party entanglement
can also be investigated.

\acknowledgments We acknowledge valuable discussions with Dr. D.
Yang. This work is supported by NSFC Grant No. 90103022, 10225419,
and 10405019.

\end{document}